\documentclass[twocolumn,         
               showpacs,longbibliography,            
               showkeys,preprintnumbers,     
               aps,                 
               prd,          	    
               letterpaper,             
               superscriptaddress,      
               nofootinbib,         
               tightenlines,        
               floats,floatfix      
               ]{revtex4-1}
\usepackage{physics}
\usepackage[utf8]{inputenc}
\usepackage[T1]{fontenc}
\usepackage{epsf}
\usepackage{graphicx,color}
\usepackage{subfigure}
\usepackage{latexsym}
\usepackage{amsmath,amssymb}        
\usepackage[colorlinks=true,linkcolor=blue,citecolor=blue]{hyperref}
\usepackage{mathrsfs}
\usepackage{comment}
\usepackage{soul}
\usepackage{listings}
\usepackage{cancel}
\definecolor{purple}{rgb}{0.58,0.0,0.83}
\definecolor{orange}{rgb}{1,0.5,0}
\DeclareSymbolFontAlphabet{\mathrsfs}{rsfs}
\DeclareMathAlphabet{\mathcal}{OMS}{cmsy}{m}{n}

\begin{document}

\title{Plain Convolution Encryption as an Alternative to Overcoming the Limitations of Synchronization-Based Methods}

\author{Flavio Rosales-Infante}
\email{flavio.rosales.infante@umich.mx }
\affiliation{Instituto de F\'{\i}sica y Matem\'{a}ticas, Universidad
              Michoacana de San Nicol\'as de Hidalgo. Edificio C-3, Cd.
              Universitaria, 58040 Morelia, Michoac\'{a}n,
              M\'{e}xico.}

\author{M.L. Romero-Amezcua}
\email{maria.romero.amezcua@umich.mx }
\affiliation{Facultad de Ciencias F\'isico Matem\'aticas, Universidad
              Michoacana de San Nicol\'as de Hidalgo. Edificio ALFA, Cd.
              Universitaria, 58040 Morelia, Michoac\'{a}n,
              M\'{e}xico.}  

\author{Iv\'an  \'Alvarez-Rios}
\email{ivan.alvarez@umich.mx}
\affiliation{Instituto de F\'{\i}sica y Matem\'{a}ticas, Universidad
              Michoacana de San Nicol\'as de Hidalgo. Edificio C-3, Cd.
              Universitaria, 58040 Morelia, Michoac\'{a}n,
              M\'{e}xico.}

\author{F. S Guzm\'an}
\email{francisco.s.guzman@umich.mx}
\affiliation{Instituto de F\'{\i}sica y Matem\'{a}ticas, Universidad
              Michoacana de San Nicol\'as de Hidalgo. Edificio C-3, Cd.
              Universitaria, 58040 Morelia, Michoac\'{a}n,
              M\'{e}xico.}


\begin{abstract}
This paper revisits the send/retrieve message process using synchronization of the Lorenz system with a monochromatic message. We analyze how the fidelity of the retrieved signal depends on the message frequency and demonstrate message hacking through Fourier spectrum analysis. Various parameters affecting fidelity and noise in the hacked signal are also examined. Additionally, we transmit text messages recovered through synchronization and investigate their vulnerability to hacking. As a countermeasure, we propose a method to send both types of messages using the convolution as the encryption function to hide the message in the chaotic signal. This approach enhances retrieval fidelity and significantly increases resistance to hacking compared to synchronization-based methods.
\end{abstract}

\maketitle


\section{Introduction}
\label{sec:introduction}

The application of chaos theory to secure communication has led to a variety of encryption methods that take advantage of the  unpredictability of chaotic systems. Early developments using chaos synchronization, demonstrated the potential for transmitting signals securely through synchronized chaotic dynamics \cite{early_use}. Building on these principles, numerous encryption techniques have been proposed, employing chaotic systems to encode and decode messages, with certain success in the encryption of images \cite{Lorenz_map, Lorenz_img}, audio \cite{audio}, and other data types.

Among the more sophisticated approaches, chaotic oscillators have been utilized to implement secure message-masking methods by generating pseudorandom keys \cite{PhysRevE.66.017202}, while chaos synchronization in coupled map lattices has been shown to achieve a balance between encryption speed and security \cite{PhysRevE.66.065202}. Time-delayed coupling mechanisms have also enabled secure key exchange protocols, reducing the probability of successful synchronization by an attacker \cite{PhysRevE.72.016214}. Advances in hyperchaotic systems, such as those described in \cite{PhysRevE.80.066209}, have improved encryption, achieving multi-user secure data transmission.

Vulnerabilities in chaos-based cryptographic schemes have also been identified. Flaws such as limited key space, insensitivity to key mismatches, and susceptibility to known-plaintext attacks have been highlighted in \cite{Li_2005}, while synchronization-based methods face risks from intruders reconstructing sender dynamics via return maps \cite{BU2004919}. Moreover, one-dimensional maps commonly used in cryptographic algorithms suffer from dynamical degradation, making them vulnerable to signal estimation techniques and other forms of attack \cite{map_vul, review, enhanced_map}. Specific weaknesses in image encryption schemes have been exploited through methods such as chosen-plaintext attacks and genetic algorithm-based decryption \cite{breaking1, breaking2, breaking3}.

This paper revisits the use of Lorenz system synchronization for message transmission. Starting with a monochromatic signal, we analyze how the fidelity of the retrieved signal depends on the frequency of the transmitted message and demonstrate that the method is susceptible to hacking via Fourier spectrum analysis. Furthermore, we extend the study to text message transmission, evaluating the recovery and security of such messages. As an alternative, we propose a method called Plain Convolution Encryption (PCE) for message encryption. We demonstrate that this method not only enhances fidelity but also provides increased resistance to decryption attempts compared to synchronization-based schemes.

The paper is organized as follows. In Section \ref{sec:encryption} we briefly describe the method to send a message and set notation, while in Section \ref{sec:preliminaries} we evaluate the effectiveness of using synchronization. In Section \ref{sec:chaotic mask} we present the results of the PCE approach and finally in Section \ref{sec:conclusions} we draw some conclusions.

\section{General transmission}
\label{sec:encryption}

{\it Message encryption.} Consider the dynamical system in the form

\begin{equation}
\left\lbrace
\begin{matrix}
    \dot{\vec{x}} & = & \vec{f}(t, \vec{x}; \vec{\alpha}) \\
    \vec{x}(0) & = & \vec{x}_0 \\
\end{matrix}
\right.
    \label{eq: DynSys}
\end{equation}

\noindent where $\vec{x} = (x_1, x_2, \ldots, x_k)$ is a vector of $k$ components $x_i$ with $i=1,2,\ldots,n$, which are real functions of time $t$, that represent the solution functions of  the system,
finally $\vec{\alpha} = (\alpha_1, \alpha_2, \ldots, \alpha_m)$ are the parameters of the system. In general, our system has the following properties:  1) $\vec{f}$ is a nonlinear function, 2) the system is at least three-dimensional, $n \geq 3$, and 3) the solution of the system $\vec{x}(t)$ is sensitive to the initial conditions $\vec{x}_0$.
With these properties, the dynamical system can exhibit chaotic behavior. Chaotic trajectories $\vec{x}(t)$ resulting from the numerical solution of system (\ref{eq: DynSys}), can be used to encrypt a message $m(t)$ difficult to decode because it is screened within the chaotic signal. The encryption process involves defining an encryption function $g$ that takes the message and the chaotic trajectory and gives the encrypted message:

\begin{eqnarray}
    m_{e}(t) = g(m(t); \vec{x}),
    \label{eq: encryption function}
\end{eqnarray}

\noindent while a requirement for this function is to be invertible in order to allow the recovery of the original message,

\begin{equation}
    m(t) = g^{-1}(m_e(t); \vec{x}).
\end{equation}

\noindent Thus, when the receiver reads the encrypted message $m_e(t)$, the only way to decrypt it is by knowing the decryption key, which is $\textbf{key}_c = \{\vec{x}_0, \vec{\alpha}, \text{model}, \text{encryption function}, \text{method}, \Delta t\}$. Here $\vec{x}_0$ and $ \vec{\alpha}$ are initial conditions and a set of parameters, the ``model'' is the particular set of equations that compose the system, ``method'' refers to the numerical method used to integrate the nonlinear system and uses resolution $\Delta t$.

{\it Chaos Synchronization for Message Transmission.} 
Chaos synchronization occurs when two systems, starting from different initial conditions, evolve to follow the same chaotic trajectory. In this work, we use complete synchronization, where the state variables of the receiver system fully match those of the emitter after a transient period. The implementation defines an emitter system, in our case the Lorenz system with certain parameters, and a receiver system that can synchronize with the emitter. The emitter adds a message embedded within one of the solution functions of the chaotic solution, and the receiver, synchronized with the emitter, recovers the message by subtracting its own chaotic trajectory, as described in \cite{Sync_Chaos}.

Specifically, if the chaotic signal from the emitter is one of the components solutions of system (\ref{eq: DynSys}) called $u(t)$, we use addition as encryption function, so that the emitted signal is $m_e(t) = u(t)+m(t)$. Later on the receiver generates its own $u_r(t)$, and recovers the message with the subtraction

\begin{equation}
	m_r(t) = m_e(t) -u_r(t). \label{recv_msg}
\end{equation}

\noindent Due to the properties of chaos synchronization, the key to decrypt the message in this case is $\textbf{key}_{cs} = \{\vec{\alpha}, \text{model}, \text{method}, \Delta t\}$, where the encryption function is omitted from the key, because in chaos synchronization the message is  added to the chaotic signal. We will show that this encryption function is not very safe nor accurate.


\section{Use of synchronization with the Lorenz system}
\label{sec:preliminaries}

\subsection{Workhorse example of sending a message}
\label{subsection:gaussian_sync}

\begin{figure}
    \centering
    \includegraphics[width=4cm]{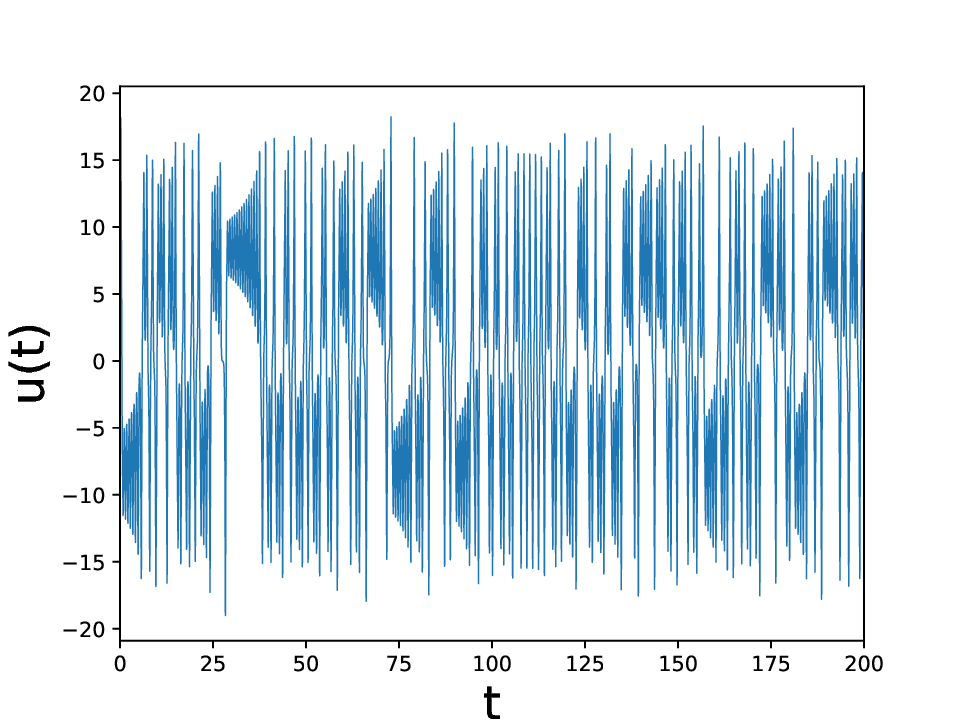}
    \includegraphics[width=4cm]{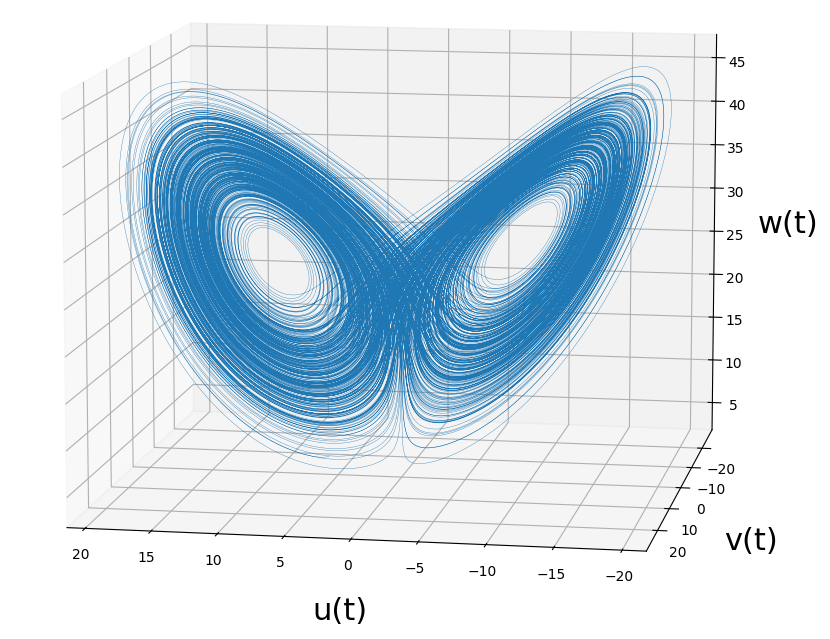}
    \caption{Solution of the Lorenz system in the chaotic regime for the initial conditions $(u_0, v_0, w_0) = (5,5,5)$ and  parameters $(a, b, r) = (10, \frac{8}{3}, 28)$. This solution is constructed using resolution $\Delta t = 0.001$ over the interval $t \in [0, 200]$ and the RK4 integrator. The time-series of $u(t)$ and the phase space trajectory of the system are shown.}
    \label{fig: LorenzSolution}
\end{figure}

The model we use is the Lorenz system:

\begin{eqnarray}
    \dot{u} & = & a(v - u),  \nonumber\\
    \dot{v} & = & ru - v - uw, \label{eq:Lorenz_system}\\
    \dot{w} & = & uv - bw, \nonumber
\end{eqnarray}

\noindent in the chaotic regime, for which we set the parameters to $(a, b, r) = (10, \frac{8}{3}, 28)$, with initial conditions $(x_0, y_0, z_0) = (5, 5, 5)$. Figure \ref{fig: LorenzSolution} shows the numerical solution of $u(t)$ in the interval $t \in [0, 200]$ with a resolution $\Delta t = 0.001$. The numerical solution is calculated using a standard RK4 method.

As an elementary example, we use for emission and recovery a monochromatic function modulated with a Gaussian:

\begin{equation}
    m(t) = A \sin(\omega t) e^{-(t - 100)^2}, \label{eq:m_gauss}
\end{equation}

\noindent of amplitude $A$ and frequency $\omega$, centered at $t=100$.

\subsection{Synchronization and message recovery}

The receiver consists of a system defined with the following modified Lorenz system:

\begin{eqnarray}
        \dot{u}_r & =& a(v_r - u_r),\nonumber\\
        \dot{v}_r & =& rm_e(t) - v_r - m_e(t)w_r,\label{eq:Lorenz_rcv}\\
        \dot{w}_r & =& m_e(t)v_r - bw_r,\nonumber
\end{eqnarray}

\noindent where the principal characteristic is that $m_e(t)$ in the right hand sides of the equations, is the solution of the emitter system plus the sent message. Synchronization allows to use a key such that $\vec{\alpha}$ contains the parameters $a,b,r$ and there is no need to include the initial conditions in the key, an advantage of the synchronization based method.  Thus, for the receiver system we use the initial conditions $(u_{r_0}, v_{r_0}, w_{r_0}) = (25, 6, 50)$, which are different from those used by the emitter since they are not in the key, while the parameters of the emitter $(a,b,r) = (10, \frac{8}{3}, 28)$ and the resolution $\Delta t = 0.001$ are part of the key, that we use to calculate the solution of (\ref{eq:Lorenz_rcv}).

\begin{figure}
    \centering
    \includegraphics[width=4cm]{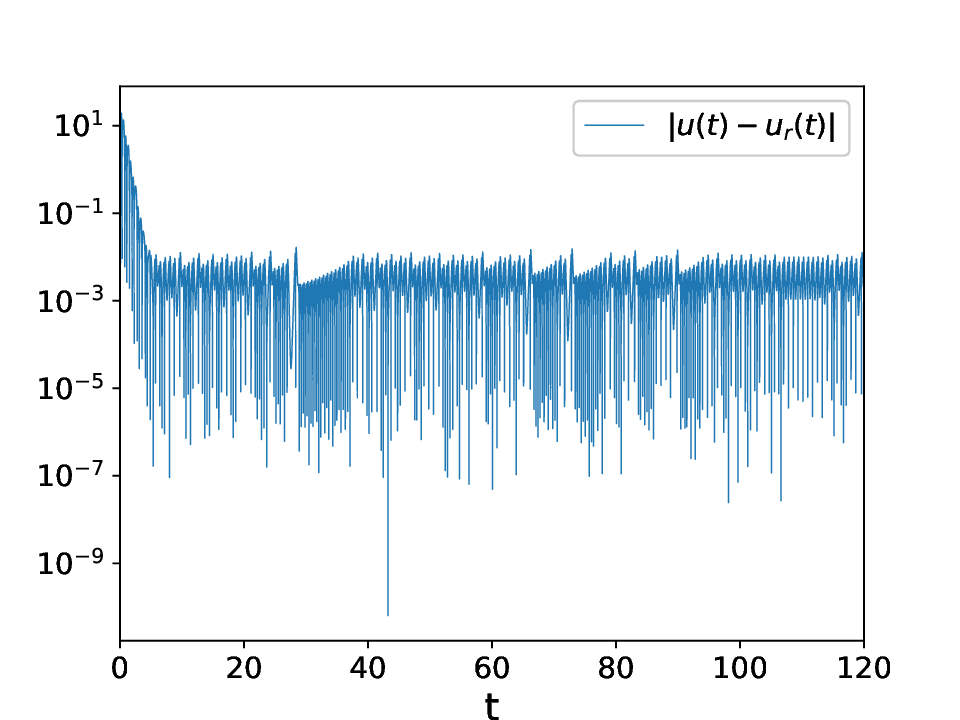}
    \includegraphics[width=4cm]{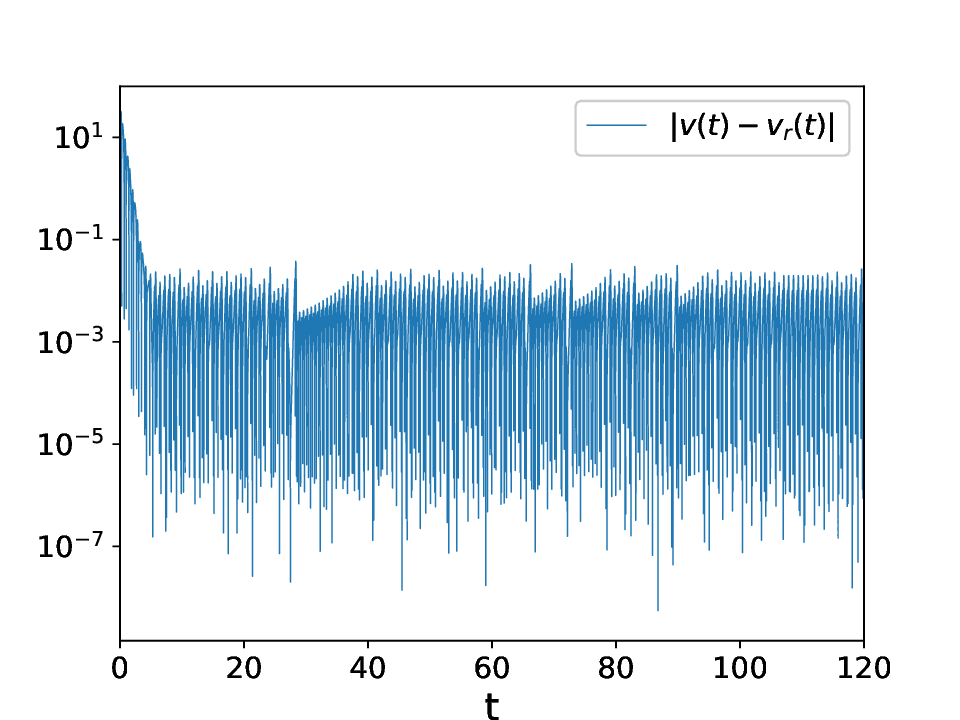}
    \includegraphics[width=4cm]{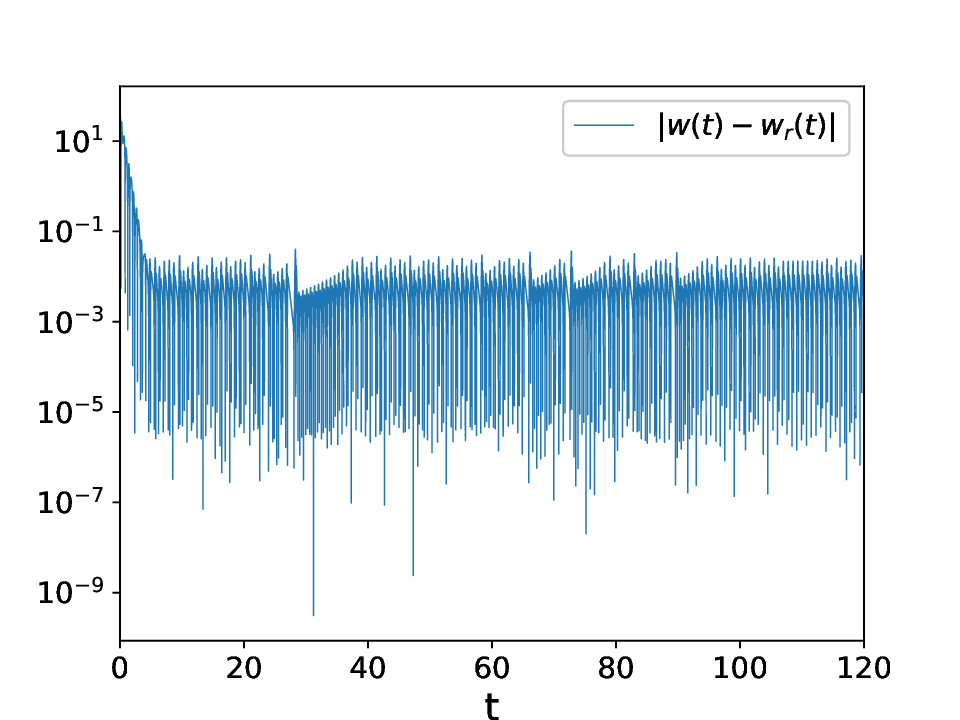}
    \caption{Time series showing the synchronization of the emitter and the receiver solutions $(u(t),v(t),w(t))$ and $(u_r(t),v_r(t),w_r(t))$ in a time window of $t\in[0,120]$. Notice that the message $m(t)$ has a Gaussian envelope centered at $t=100$, a way after the synchronization of the system happens.}
    \label{fig: Synchronization}
\end{figure}

In order to calibrate the numerical solution of emitter and receiver, we test that synchronization happens without the encrypted message, thus the emitted signal is only the solution function $u$ (with $m(t)=0$) of the Lorenz system (\ref{eq:Lorenz_system}). As the time evolves, the solution of the receiver $u_{r}, v_{r}, w_{r}$, start to tend toward the solution of the emitter $u, v, w$ as shown Figure \ref{fig: Synchronization}. When this happens, the two systems are said to be synchronized.

Now that we know that synchronization happens, we send the message in a time window where the systems are synchronized using the encryption function $m_e(t)=u(t) + m(t)$. With this signal $m_e(t)$, the solutions $u_{r}, v_{r}, w_{r}$ of the receiver are calculated, which, as seen before, tend to $u, v, w$. Then the message $m_r(t)$ can be recovered using the subtraction in \eqref{recv_msg}.

\begin{figure}
    \centering
    \includegraphics[width=8cm]{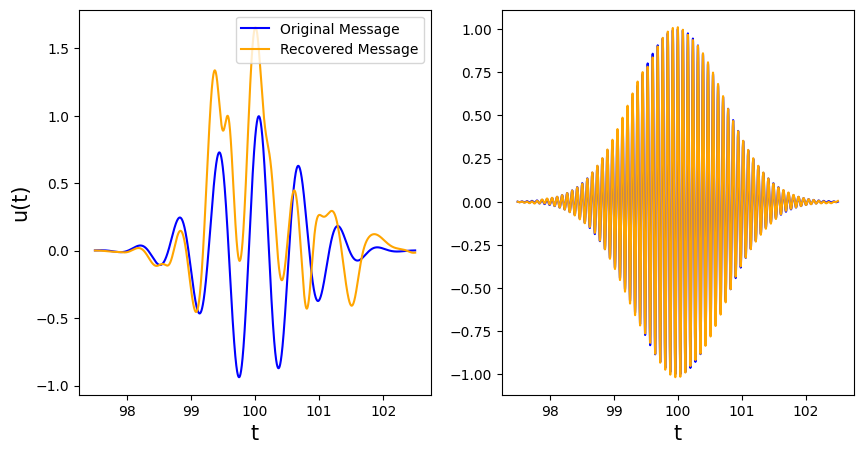}
    \caption{Comparison between $m(t)$ and $m_r(t)$ for $A = 1$ and two different frequencies, $\omega = 10$ (left) and $\omega = 80$ (right).}
    \label{fig: Messages}
\end{figure}

\begin{figure}
    \centering
    \includegraphics[width=8cm]{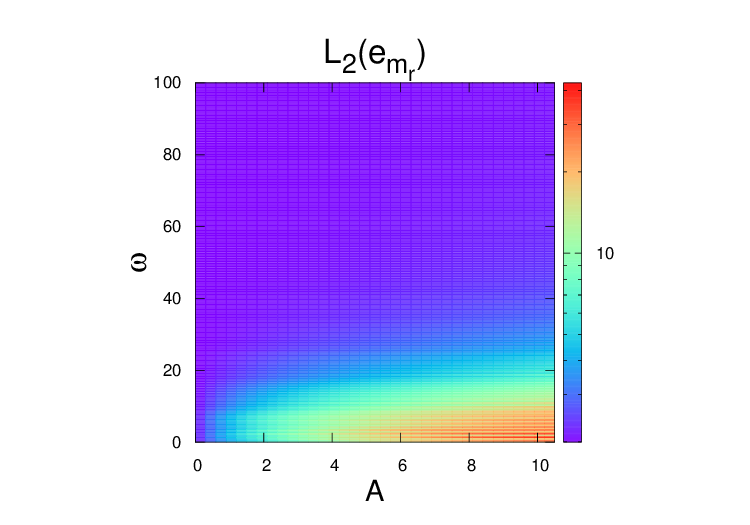}
    \caption{The $L_2$ norm of $e = m(t) - m_r(t)$ for a set of messages $m(t)$ of type (\ref{eq:m_gauss}) with parameters in the range $A\in[0.1,11]$ and $\omega \in[0.1,100]$. This plot illustrates how the fidelity in the message recovery degrades for small frequencies and big amplitudes.}.
    \label{fig: Norms}
\end{figure}

{\it Fidelity and error.} The message reconstructed by the receiver $m_r(t)$ is not exactly the original one $m(t)$.The differences between these two are showen in Figure \ref{fig: Messages} for two values of $\omega$. At first sight one can guess that the differences are bigger/smaller for small/high frequencies. 
In order to illustrate how fidelity depends on $A$ and $\omega$, we calculate the $L_2$ norm of the error in the recovered signal $e = m(t) - m_r(t)$, which is summarized in Figure \ref{fig: Norms}. The errors are bigger for small frequencies and big amplitudes of this monochromatic message.
This is a considerably simple message, but helps illustrating how fidelity degrades in terms of the properties of the signal itself. However, this simple case reflects the quality of $m_r(t)$.

\begin{figure}
	\centering
	\includegraphics[width=8cm]{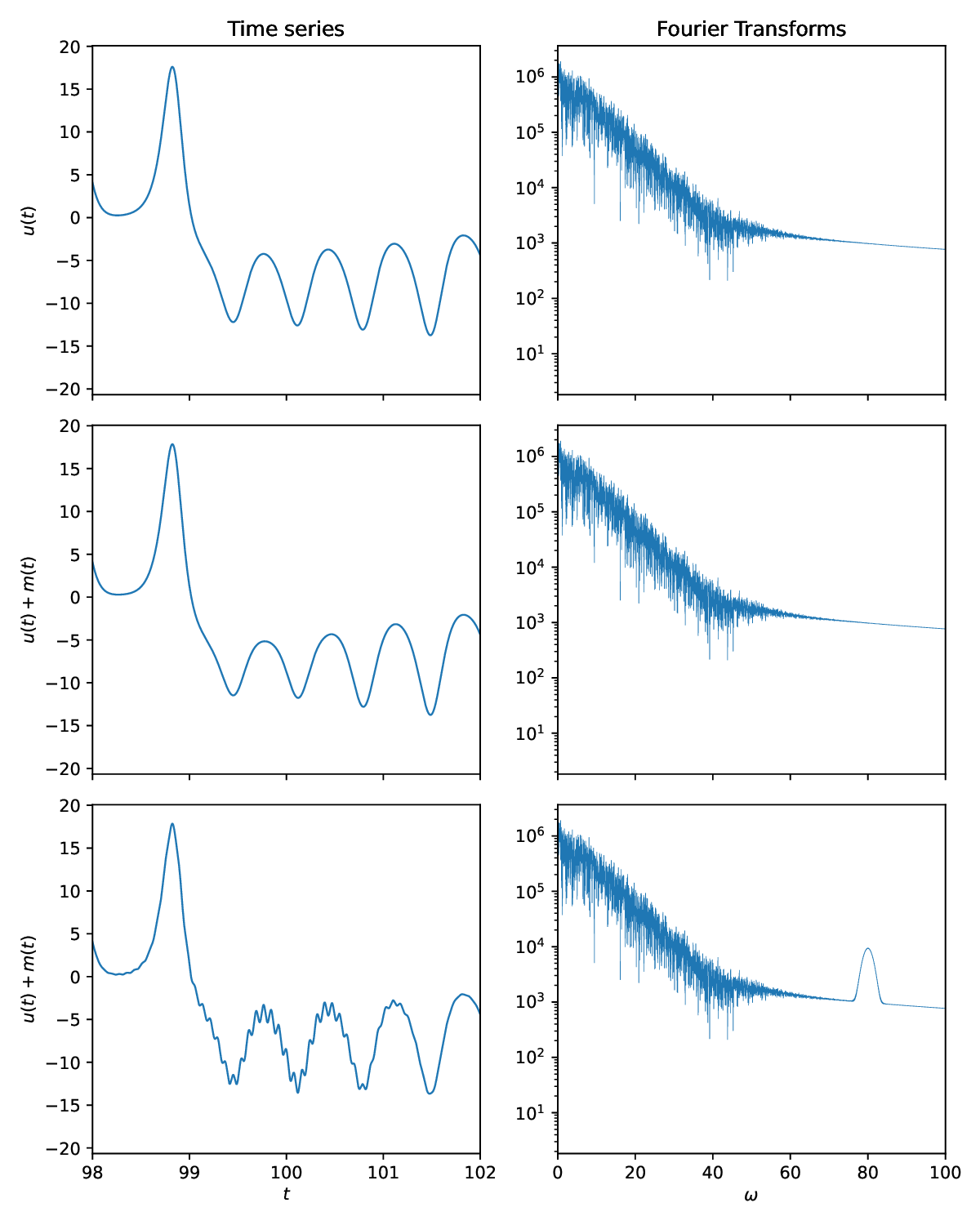} 
	\caption{\label{fig: FT_gauss} Time series of $u(t)+m(t)$ along with its Fourier Transform of three cases, the pure chaotic signal case $m(t)=0$ (top), the low frequency message (\ref{eq:m_gauss}) and $\omega=10$ (middle) and the high frequency message with $\omega=80$ (bottom), both with $A=1$. }
\end{figure}

Figure \ref{fig: FT_gauss} shows the signal of $u(t)$ and $m_e(t)=u(t)+m(t)$ in a neighborhood where the Gaussian is maximum for $\omega=10$ and 80. Additionally, the corresponding Fourier Transform (FT) calculated in the whole time domain of the solution is included. For the control case with $m(t)=0$, the solution of the Lorenz system along with its FT is shown, which indicates that the signal has more power in the low frequency domain.
The second and third rows of Figure \ref{fig: FT_gauss} show the signal and FT of the cases with $\omega=10$ and $\omega=80$, respectively. In the later case a high frequency mode can be seen in time and frequency domains, revealing that there is something anomalous in the time-series, possibly a signal.

Here is the dilema. A message sent with small frequencies hides very well the message within the chaotic signal, as seen in the Fourier Spectra of Figure \ref{fig: FT_gauss}, unfortunately the fidelity of $m_r(t)$ is not very good as illustrated in Figure \ref{fig: Messages}. On the other hand, the fidelity of the recovered message is good when using high frequencies as seen in Figure \ref{fig: Messages}, however, a FT of the signal reveals a clearly identifiable glitch in the tail region of the Fourier Spectrum in Figure \ref{fig: FT_gauss}, that can be associated to a message.

\subsection{Hacking this simple signal} 
\label{sec: hacking}

If a feature within the high frequency part of the spectrum is detected, the message can be decrypted without needing the $\mathbf{key}_{cs}$. The first step is to view the transmitted signal in the Fourier space. Since the message can be identified with a glitch in the tail zone, the message modes can be clearly detected even if they cannot be seen in the time series, as shown in Figure \ref{fig: FT_gauss} for the case with $\omega = 80$. \\

\noindent The second step is to remove frequencies similar to those of the solution $u(t)$. This is achieved using a high-pass filter with a cutoff frequency $\omega_{co}$, beyond which the message modes should be present. The third step is to apply the inverse Fourier transform to the filtered signal to retrieve the message as follows  

\begin{eqnarray}
        m_h(t) &=& \mathcal{F}^{-1} \left[HPF(\mathcal{F}\left[u_r(t)\right])\right], \label{eq:m_hacked}\\ 
        HPF(f(\omega)) &=& \begin{cases}
            0, \quad if\, |\omega|< \omega_{co} \\
            f(\omega), \quad if \, |\omega|\geq \omega_{co},            
        \end{cases} \label{eq:HPF} 
\end{eqnarray}

\noindent where $m_h(t)$ is the hacked message, and $HPF$ a high-pass filter.

\begin{figure}
    \centering
    \includegraphics[width=8cm]{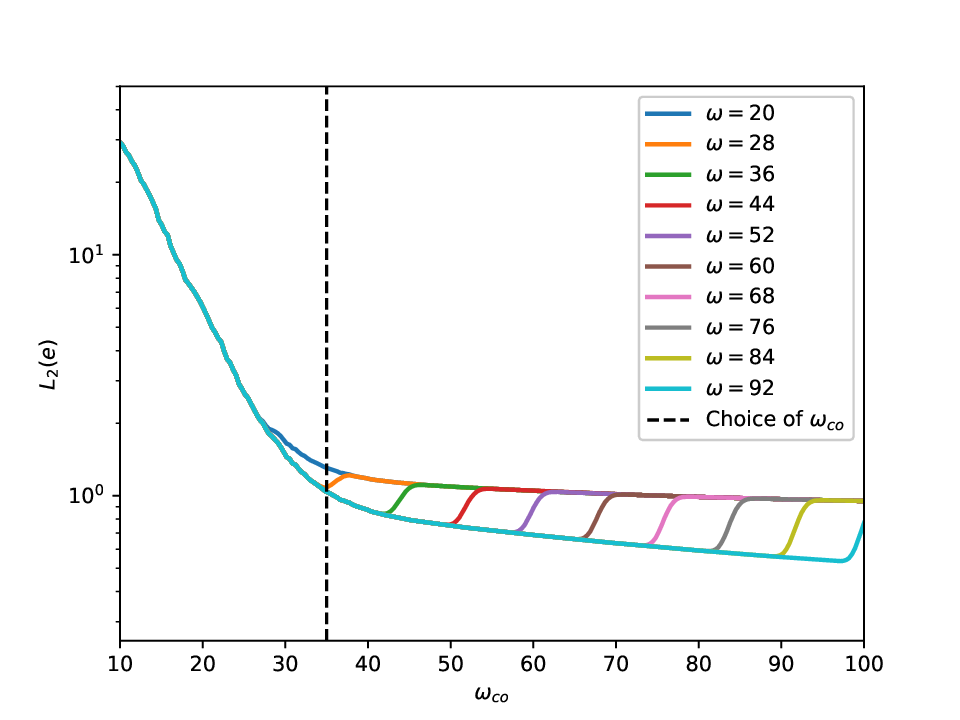}
    \caption{$L_2$ norm of the error $e_h = m_h(t)-m(t)$ for a set of messages with different frequencies and the same amplitude $A=1$, the vertical black line shows election of $\omega_{co}$.This graph demonstrates the impact of cutoff frequency for $m_h(t)$ as well as the limitation of this method at low frequencies.}
    \label{fig:Omega_cutoff}
\end{figure}

\noindent The accuracy of the hacked message depends on the choice of the cut-off frequency $\omega_{co}$, a bigger value will give high accuracy, but will loose message information in the low frequency part of the spectrum. For this monochromatic example, the $L_2$ norm of the error between the message sent and the hacked message, $e_h = m_h(t) - m(t)$, was measured using cutoff frequencies using different values of $\omega_{co}$ between $[0,100]$ and the result is in Figure \ref{fig:Omega_cutoff}. All sent messages share the same error for low frequencies, but, as $\omega_{co}$ approaches the message frequency $\omega$, the error increases until it becomes constant, as the message is eliminated along with  the system's solution. Therefore, a good election is the interval $\omega_{co} \in [35,40]$ because is the region where the interference of the system's solution can be avoided and have a considerable fidelity in the hacked message.

Up to this point we have illustrated how to send, receive and hack a monochromatic message. Thus we show next that more complex messages can also be sent and recovered with synchronization.

\subsection{Sending, receiving and hacking text messages}
\label{subsection:text_sync}

In this section, a text message is translated, loaded, sent, and decrypted using the synchronization of chaos. For this, the following famous text is used:

\begin{quote}
``Many years later, as he faced the firing squad, Colonel Aureliano Buend\'ia was to remember that distant afternoon when his father took him to discover ice.'' \\
- Gabriel Garc\'ia M\'arquez, \textit{Cien a\~nos de soledad}
\end{quote}

\noindent {\it Translation of the message.} Each character is converted into an integer using Unicode format according with Table \ref{Tab: LettersToAscii}, arranged as a vector of numbers $\vec{m}$. In order to hide the message within $u(t)$, we normalize the components of the message as $m_i \rightarrow m_i/300$ to have entries of order one. Then the vector with the message is defined by the following time-series:

\begin{eqnarray}
    m(t) &=& m_i \chi \left(t_0 + i n \Delta t, t_0 + (i+1)n\Delta t\right) \label{eq:msg_text(t)} \\
    \chi(a,b) &=& \begin{cases}
        1, \quad if \quad t \in \left[a,b\right) \\
        0, \quad else
    \end{cases}
\end{eqnarray}

\noindent where $\Delta t$ is the numerical resolution used to calculate $u(t)$, $t_0$ is the time when the message starts within the time series of $u(t)$, we use $n$ as the number of time steps which separate each character, and finally the label take on values $i = 0, 1, \ldots, number\,of\,characters -1$. 

{\it Loading the message.} Once $m(t)$ is generated, the signal is injected within the sender solution $u(t)$ using an Encryption Function. As seen above, there are two parameters used to inject the message, $t_0$, that for our particular text message is set to $t_0=100$, far beyond synchronization as seen in Fig. \ref{fig: Synchronization}, and 
$n$, that is related to the frequency in which the pieces of the message are sent, a small/big $n$ will correspond to high/low frequencies, that could be captured an FT like in Figure \ref{fig: FT_gauss} for the monochromatic message.

{\it Retrieving the message.} Likewise in the monochromatic signal, the message is obtained by subtracting the synchronized solution calculated by the receiver and obtaining $m_r$ using Eq. (\ref{recv_msg}). Later on we un-normalize defining $m_r \rightarrow 300m_r$ and then round each $m_r(t_i)$ to an integer in order to identify back a number with an associated character of Table \ref{Tab: LettersToAscii}.

\begin{figure}
    \centering
    \includegraphics[width=8cm]{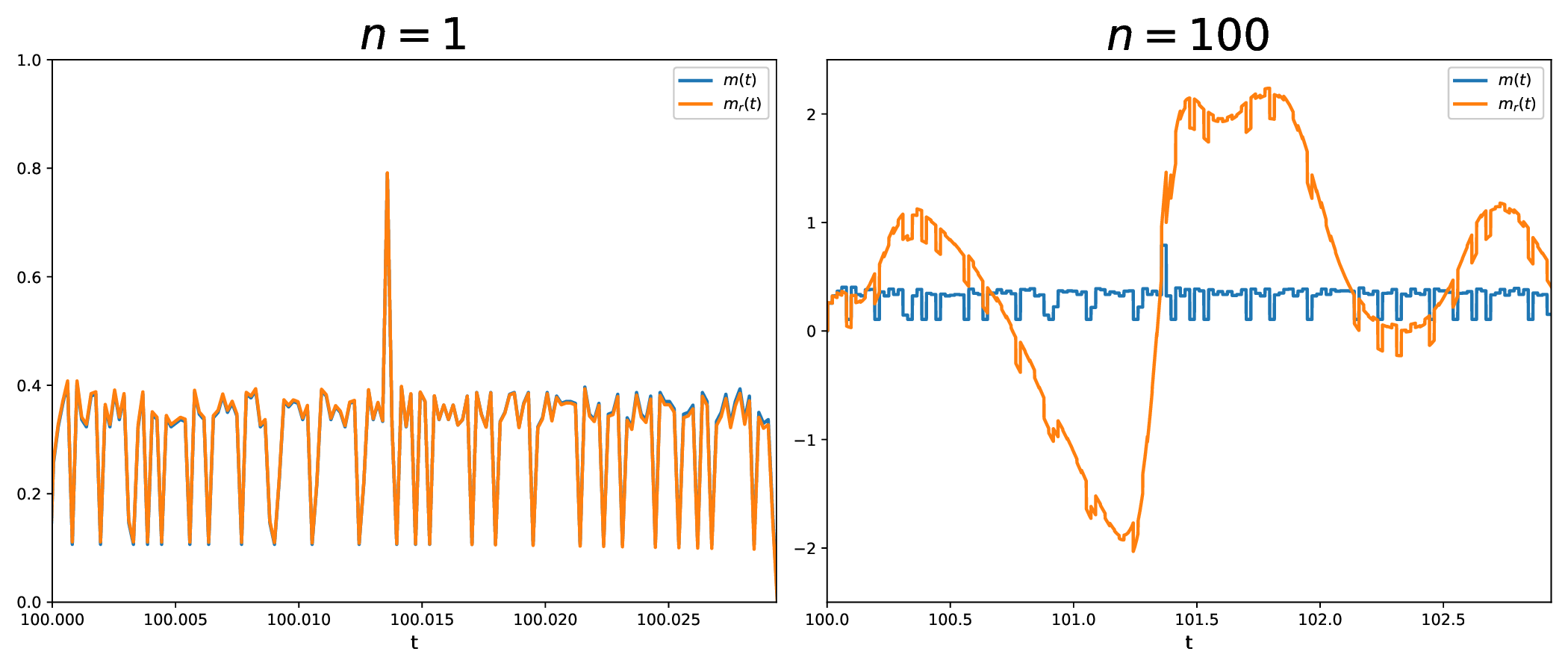}
    \caption{Comparison of $m(t)$ and $m_r(t)$ for $n=1$ (left) and $n=100$ (right). The plots show how for large $n$ the difference is considerable. The time window in this plot starts at $t_0$ and ends at the end of the message.}
    \label{fig:Sync_text}
\end{figure}

In Figure \ref{fig:Sync_text}, we shown how the receiver can unpack the message for two values of $n$, noticing that a bigger value has smaller fidelity and viceversa. The text recovered for $n=1$ is:

\begin{quote}
Nboz!zfbst!mbufs-!bt!if!gbdfe!uif!gjsjoh!trvbe- !Dpmpofm!Bvsfmjbop!Cvfod\'ia was to remember that distans\textasciigrave  esdqmnnmvgdmghre\textasciigrave sfcprmmifgkrmbg qamsbof\textasciigrave b+
\end{quote}

\noindent and now for $n=100$:

\begin{quote}
NainbNPmK\textperiodcentered \'{I}\={A}\v{D}Ĳ\'{y}\^{u}l\textperiodcentered \H{o}\'{u}\L i\ss \v{E}\^{c}\'{y}\~{n}\`{a}\"{I}³Exh\textless\textbackslash) \^{g}\^{I}\u{u} \v{z}\u{q}\v{r}\v{u}\u{m}\v{v}\u{p}\u{a}\v{l} \v{r}\u{l}\v{w}\v{v}\u{l}\u{y}\v{t}\u{s}\u{L}\v{C}\u{U}\textdegree  zX9\textquoteleft :G\textbackslash  qB\textquotedbl É\`{i}\textdegree \={i}\v{L}\'{y}o\={r} \v{s}\u{O}\v{C}\v{o}İİ\`{E}\u{A}êØ
\end{quote}

\noindent where only a few chains of characters make sense. In both cases, chaos synchronization cannot recover the message with enough fidelity to read the entire text. The percentage of fidelity for $n=1$ is $ 20.2\% $  and for $n=100$ is $0.65\%$ in this particular example.

{\it Hacking the message.} For this purpose  we use the same method as for the monochromatic signal  \eqref{eq:m_hacked},  with a cut-off frequency $\omega_{co} = 35$. The hacked message when using  $n=1$ reads

\begin{quote}
Nboz!zfbst!mbufs-!bt!ie faced the firing squad, Colonel Auqdkh`mnAtdmc\`{i}`v`rsnqdld ladqsg`schrs`msafternoon when his father took him to!ejtdpwfs!jdf
\end{quote}

\noindent whereas for $n=100$ the text is

\begin{quote}
~\&v[Saba[tk] 2u2wp(jccee"wmj\%jjqehl mtci4) Mxtuqfk=qm\textasciigrave gdii :l[cY\'{a}UkVimjqfohpdfqoXj[ bosdt|*)hjs\_ g\_ [ mdhw.yz)g[fUO]jnwz3|\{z's hUYdVepcu'uvP
\end{quote}

\noindent Notice that for $n=1$ the message is partially hacked with a percentage fidelity of $43.23\%$, because it is transmitted in a high frequency band, whereas for $n=100$ the intercepted message has $1.94\%$. A curious results, not to say ridiculous, is that  the hacked message has smaller errors than the message obtained by the synchronized receiver.

This example shows that also non-trivial messages can be hacked when using chaos synchronization, with limited -but quantifiable- fidelity in terms of the frequency band chosen for transmission. 
In summary, chaos synchronization used to send a text message is weak because recovery is difficult and it can be partially hacked.

\section{Sending a message using Plain Convolution Encryption}
\label{sec:chaotic mask}

In this section, we present the method of PCE for message encryption, with the purpose of addressing the vulnerability at high frequencies in the Fourier spectrum. Unlike in the synchronization based method, the key includes $\vec{x}_0$, initial conditions $(u_0,v_0,w_0)$ used to produce the solution of the emitter. This extended key enhances the quality of the recovered message, compared to that in the synchronization based method at low frequencies. In this way, the weight of the security is now on the Encryption Function. 

The key is then $\textbf{key}_c = \{\vec{x}_0, \vec{\alpha}, \text{model}, \text{encryption function}, \text{method}, \Delta t\}$. We illustrate the method with the Lorenz system (\ref{eq:Lorenz_system}) like before, with parameters $(a, b, r) = (10, \frac{8}{3}, 28)$, and  initial  conditions $(u_0, v_0, w_0) = (5,5,5)$. The numerical solution $u(t)$ of the emitter is calculated  in the interval $t \in [0,600]$, using resolution of $\Delta t = 0.0001$ and the RK4 integrator. 

For the encryption we use two different functions. The first one is the addition:

\begin{equation}
    m_e(t) = m(t) + u(t), \label{eq:EF_sum}
\end{equation}

\noindent and the second function is the convolution:

\begin{equation}
    m_e(t) = u(t) * m(t) = \mathcal{F}^{-1}\left[\mathcal{F}\left( u(t)\right)  \mathcal{F}\left(m(t) \right) \right]. \label{eq:EF_conv}
\end{equation}

\noindent In each case the message is retrieved by the receiver using

\begin{eqnarray}
	m_r(t) &=& m_e(t) - u(t), \label{eq:EF_inv_sum} \\ 
	m_r(t) &=& \mathcal{F}^{-1} \left[ \frac{\mathcal{F}\left( m_e(t) \right)}{\mathcal{F}\left( u(t) \right)} \right] \label{eq:EF_inv_conv},
\end{eqnarray}

\noindent  For these two functions we test fidelity and security.

\subsection{Transmission of a monochromatic wave}

We again send the monochromatic signal \eqref{eq:m_gauss}, encrypted in the chaotic signal $u(t)$ generated with the $\mathbf{key}_c$. 

{\it Retrieving the message.} For this we apply the inverse encryption functions \eqref{eq:EF_inv_sum} or \eqref{eq:EF_inv_conv}, to the encrypted message $m_e(t)$, keeping in mind that the $u_r(t)$ and $u(t)$ are practically the same, because they are generated with the same initial conditions. We present the results for each encryption function separately.

\subsubsection{Addition as encryption function}

\begin{figure}
    \centering
    \includegraphics[width=8cm]{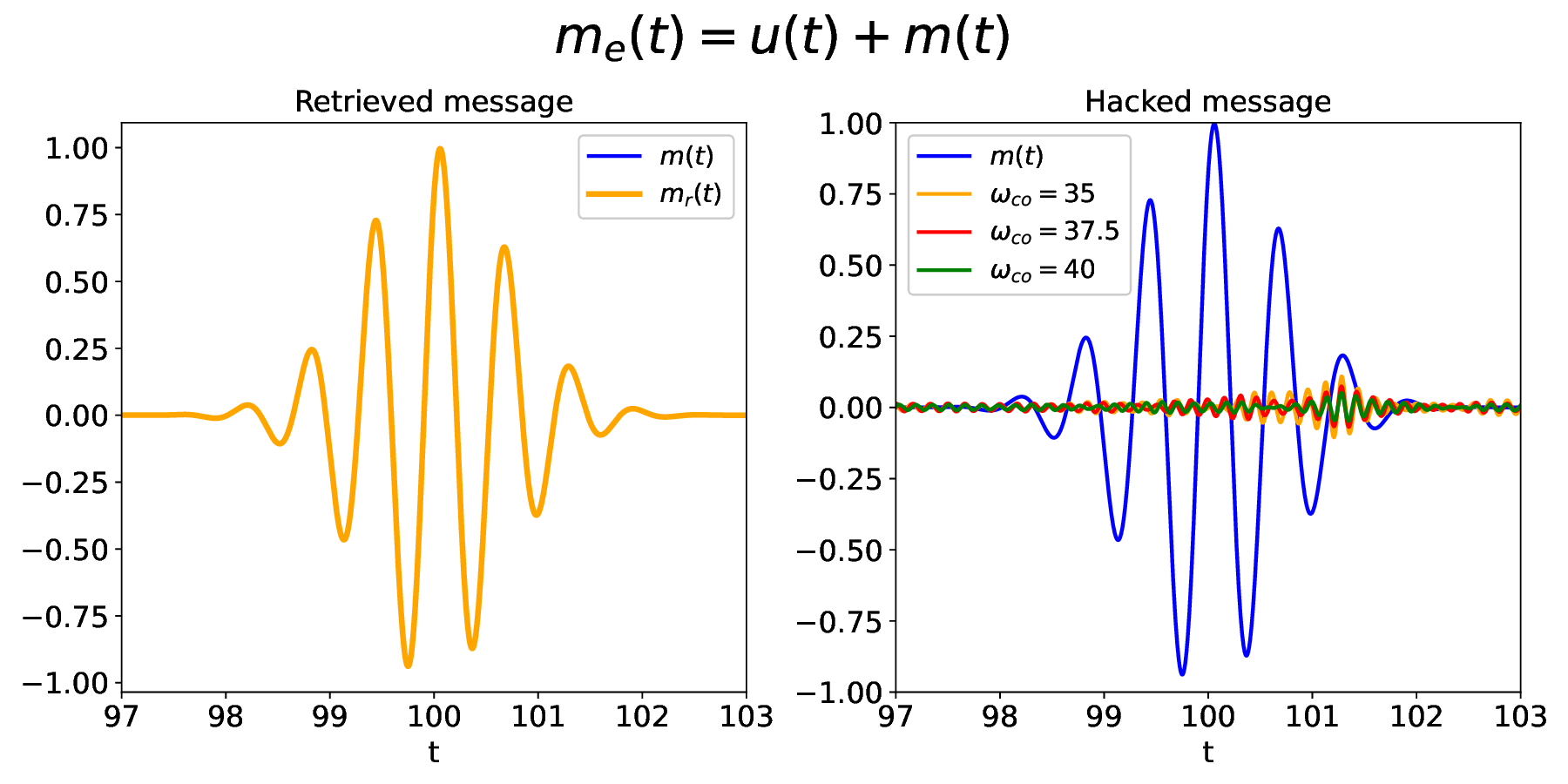}
    \includegraphics[width=8cm]{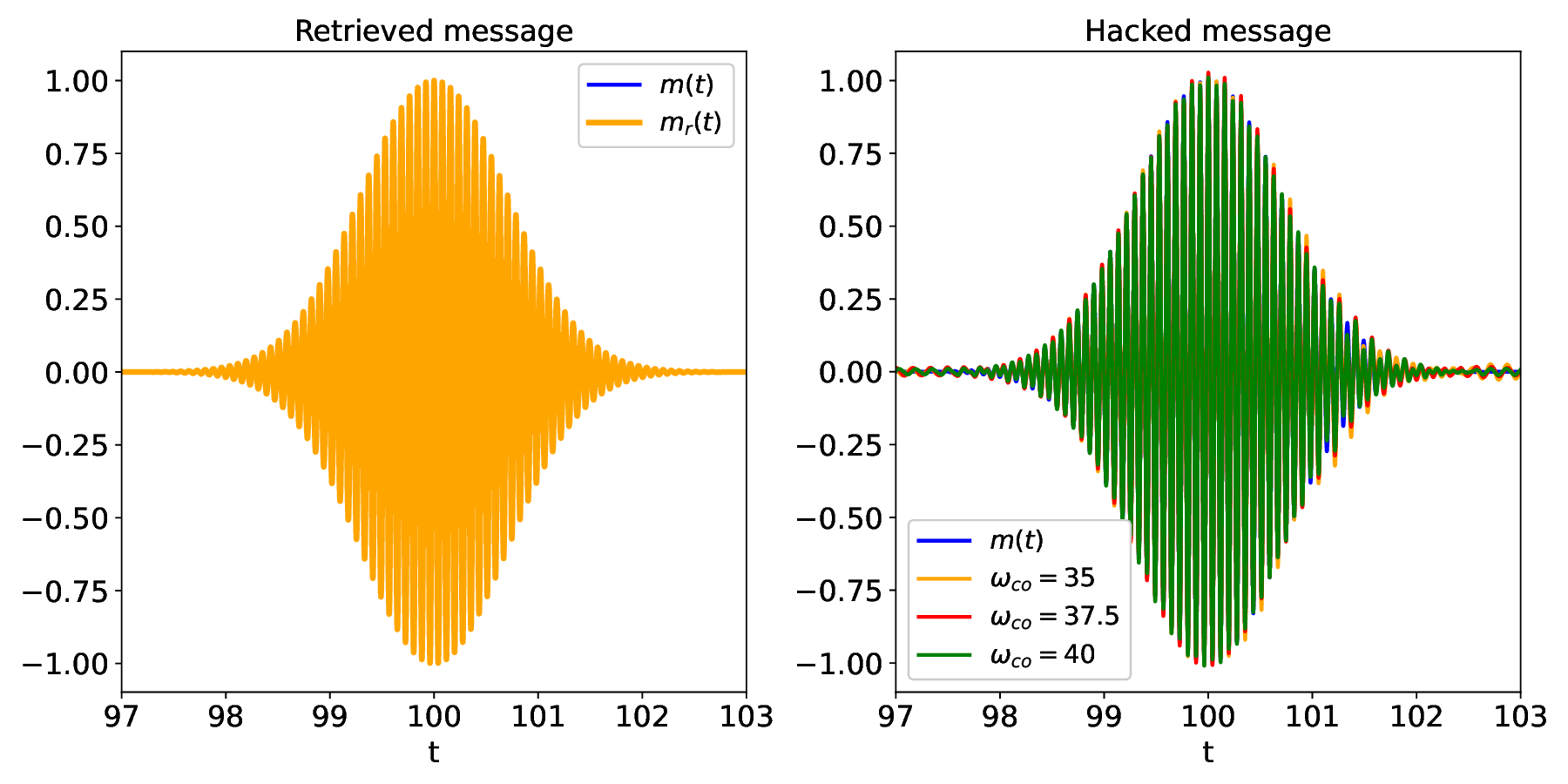}
    \caption{ Comparision of $m(t)$ and $m_r(t)$ (left), and $m(t)$ and $m_h(t)$ using varios cut-off frequencies (right), using the encryption function \eqref{eq:EF_sum}. The sent messages are monochromatic signals \eqref{eq:m_gauss} with $A=1$ and two frequencies $\omega=10$ (top) and $\omega = 80$ (bottom).}
    \label{fig:msg_gauss_sum}
\end{figure}

\begin{figure}
    \centering
    \includegraphics[width=8cm]{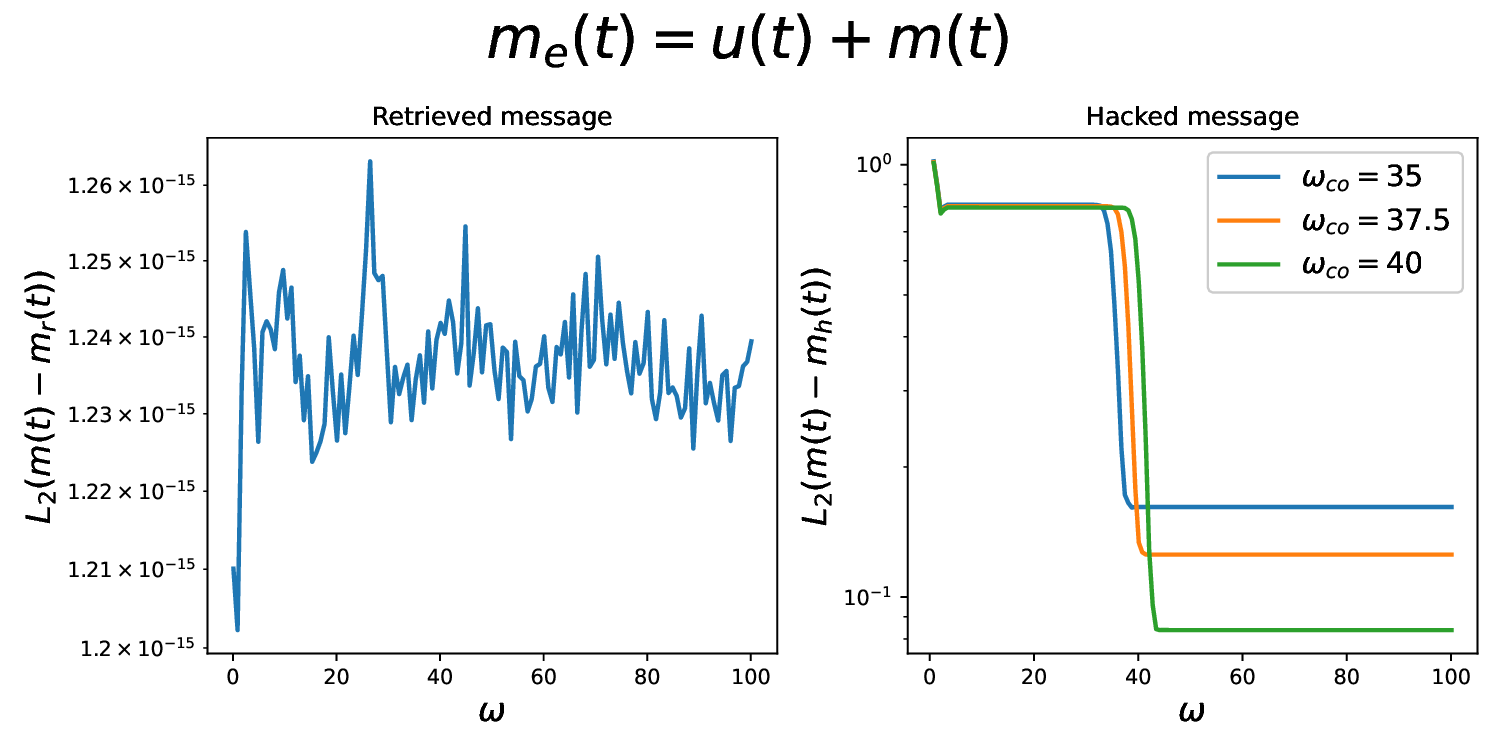}
    \caption{$L_2$ norm of $e(t)$ (left) and  $e_h(t)$ (right). Messages were encrypted using \eqref{eq:EF_sum}. The sent message $m(t)$ is a monochromatic signal defined by  \eqref{eq:m_gauss}, with $A=1$ and $\omega \in [0.1,100]$.We measure the quality of the hacked message using the method \eqref{eq:m_hacked} as function of $\omega$ with three different values of $\omega_{co}$.}
    \label{fig:error_sum}
\end{figure}

We transmit two monochromatic signals using (\ref{eq:EF_sum}) with amplitude $A=1$ and frequencies $\omega=10$ and $\omega=80$ and the recovered message can be seen in Figure \ref{fig:msg_gauss_sum}. Since the key contains the initial conditions the retrieved message $m_r(t)$ is practically the same as the sent one $m(t)$. We calculate the $L_2$ norm of $e(t)=m_r(t) - m(t)$ for $\omega \in [0.1, 100]$, which is rather insensitive to the frequency as seen in Figure \ref{fig:error_sum}.

{\it Hacking the Message.} Figure \ref{fig:msg_gauss_sum} shows the results of applying the hacking method \eqref{eq:m_hacked} using cutoff frequencies $\omega_{co} = 35, 37.5,$ and $40$. As discussed in the previous section, the low-frequency message proves more challenging to hack with high accuracy, whereas the high-frequency message is intercepted more effectively. The quality of the intercepted message is quantified by the 
$L_2$ norm of $e_h$ in Figure \ref{fig:error_sum}, which reinforces the earlier observation: low-frequency messages are harder to intercept with high fidelity.

\subsubsection{Convolution as encryption function}

\begin{figure}
    \centering
    \includegraphics[width=8cm]{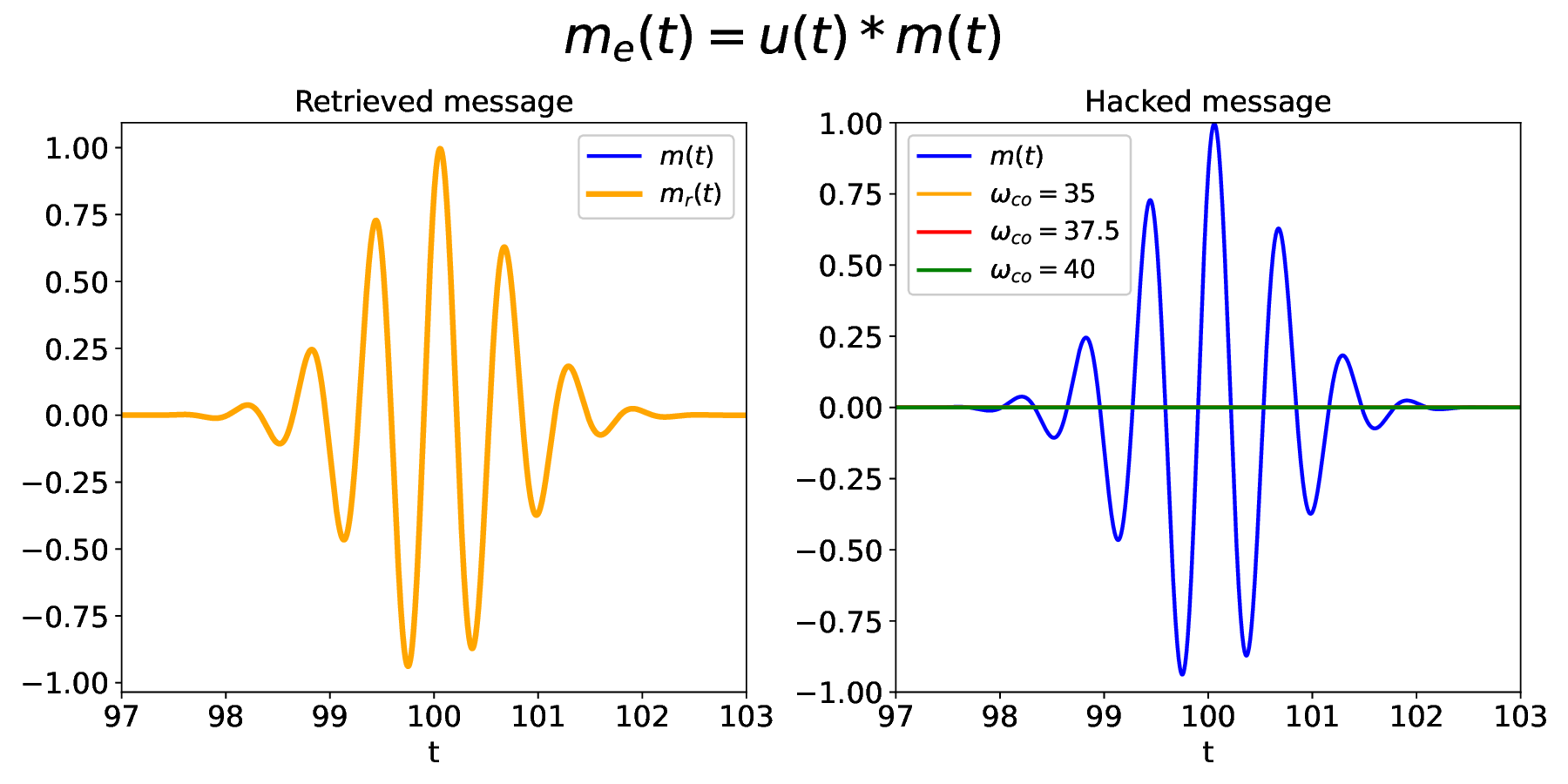}
    \includegraphics[width=8cm]{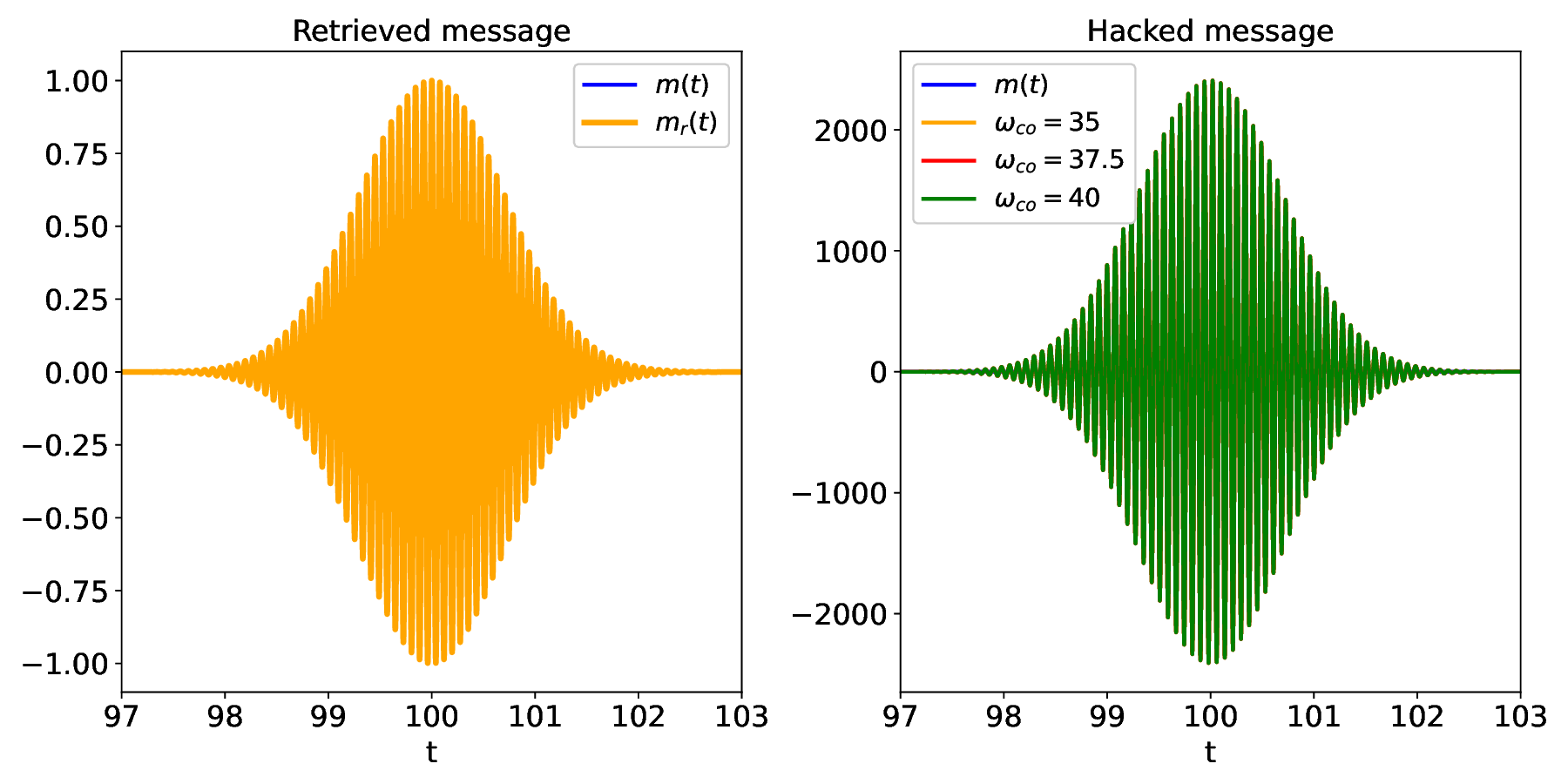}
    \caption{At the left the retrieved message $m_r(t)$ and at the right the hacked message $m_h(t)$ using various cut-off frequencies. At the top $\omega=10$ and at the bottom $\omega=80$. In this case the message is encrypted with the function (\ref{eq:EF_conv}) and recovered with  Eq. (\ref{eq:EF_inv_conv}).}
    \label{fig:msg_gauss_conv}
\end{figure}

\begin{figure}
	\centering 
    \includegraphics[width=8cm]{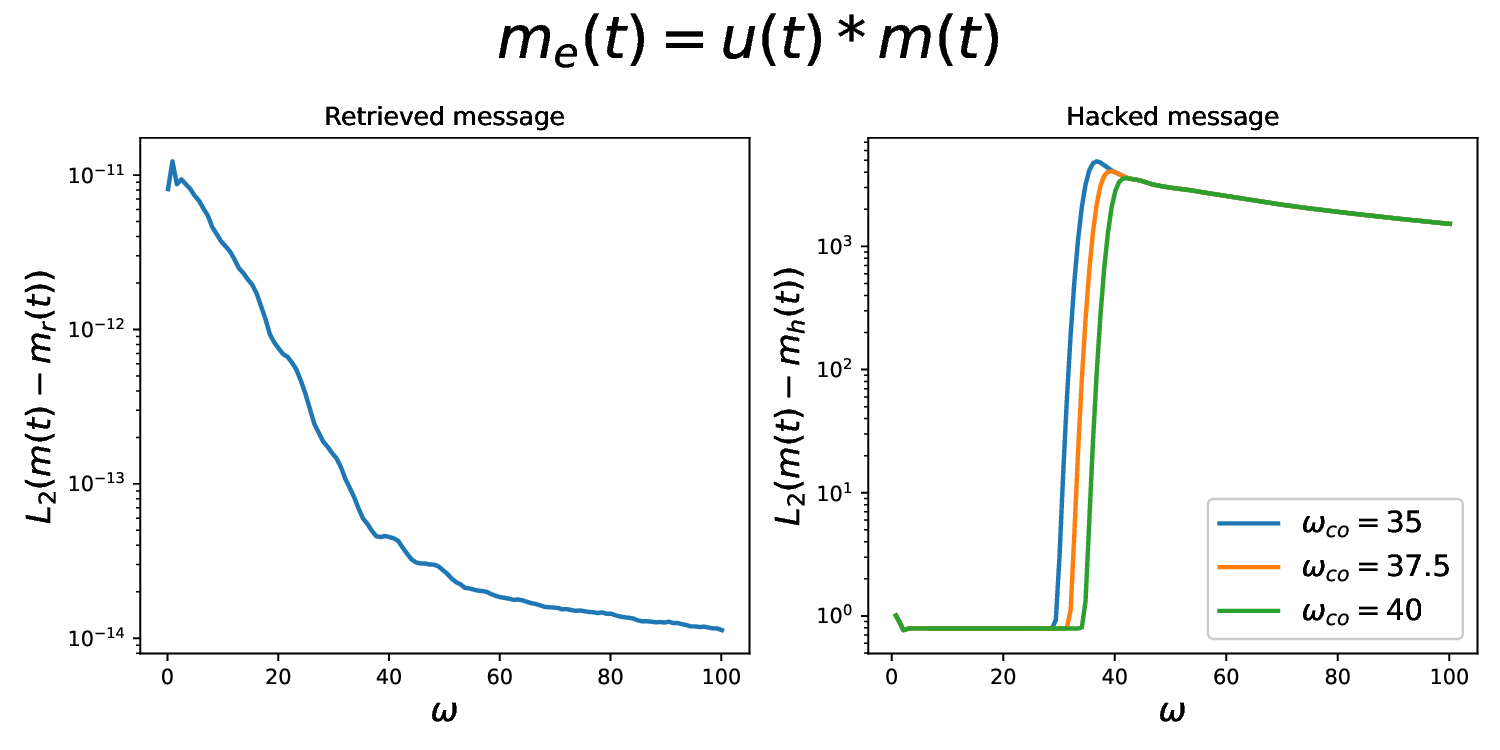}
    \caption{We show the $L_2$ norm of $e_r$ (left) and $e_h$ (right) as function of $\omega$, when the encription function is Eq. (\ref{eq:EF_conv}), hacked messages were calulated using three different values of $\omega_{co}$. These results indicate that fidelity os good in the whole range of frequencies, while the quality of the hacked signal worsens for high frequencies. The later is a behavior different from that using synchronization, where the signal was easy to hack in the high frequency region}
    \label{fig:error_convolution}
\end{figure}

We transmit now the same monochromatic messages, however this time using (\ref{eq:EF_conv}) as the encryption function. The  message $m_r(t)$ is recovered using Eq. (\ref{eq:EF_inv_conv}) and shows high quality as seen in the left column of Figure \ref{fig:msg_gauss_conv} for both frequencies  $\omega=10$ and $\omega=80$. The quality of $m_r(t)$ as function of the frequency is monitored with the $L_2$ norm of the error  in Fig. \ref{fig:error_convolution}, showing that the quality improves with frequency, although the norm of the error is or round-off error in the whole range.

{\it Hacking the message.} We attempted to hack the message and the result shown in Fig. \ref{fig:error_convolution}, where the error of $m_h(t)$ is comparable to that using synchronization for low frequencies, however it is two orders of magnitude bigger for high frequencies. For higher frequencies the message is amplified, $m_h(t)$ has the appropriate functional form but not the correct amplitude, as seen in Figure \ref{fig:msg_gauss_conv}.

These results can be summarized as follows.  Using PCE enhances fidelity compared to the use of synchronization. For monochromatic messages, fidelity improves with the extended key: however, since the encryption method remains the same, the security issues persist if we use the encryption function \eqref{eq:EF_sum}. However, the encryption function \eqref{eq:EF_conv} enhances security by preventing complete recovery of the message from the hacked signal.


\subsection{Sending a text message}

We now test how PCE works for a text message. We encrypt the same message in the same way as before, and the message is encrypted using the functions (\ref{eq:EF_sum}) and (\ref{eq:EF_conv}).

\begin{figure}
    \centering
    \includegraphics[width=8cm]{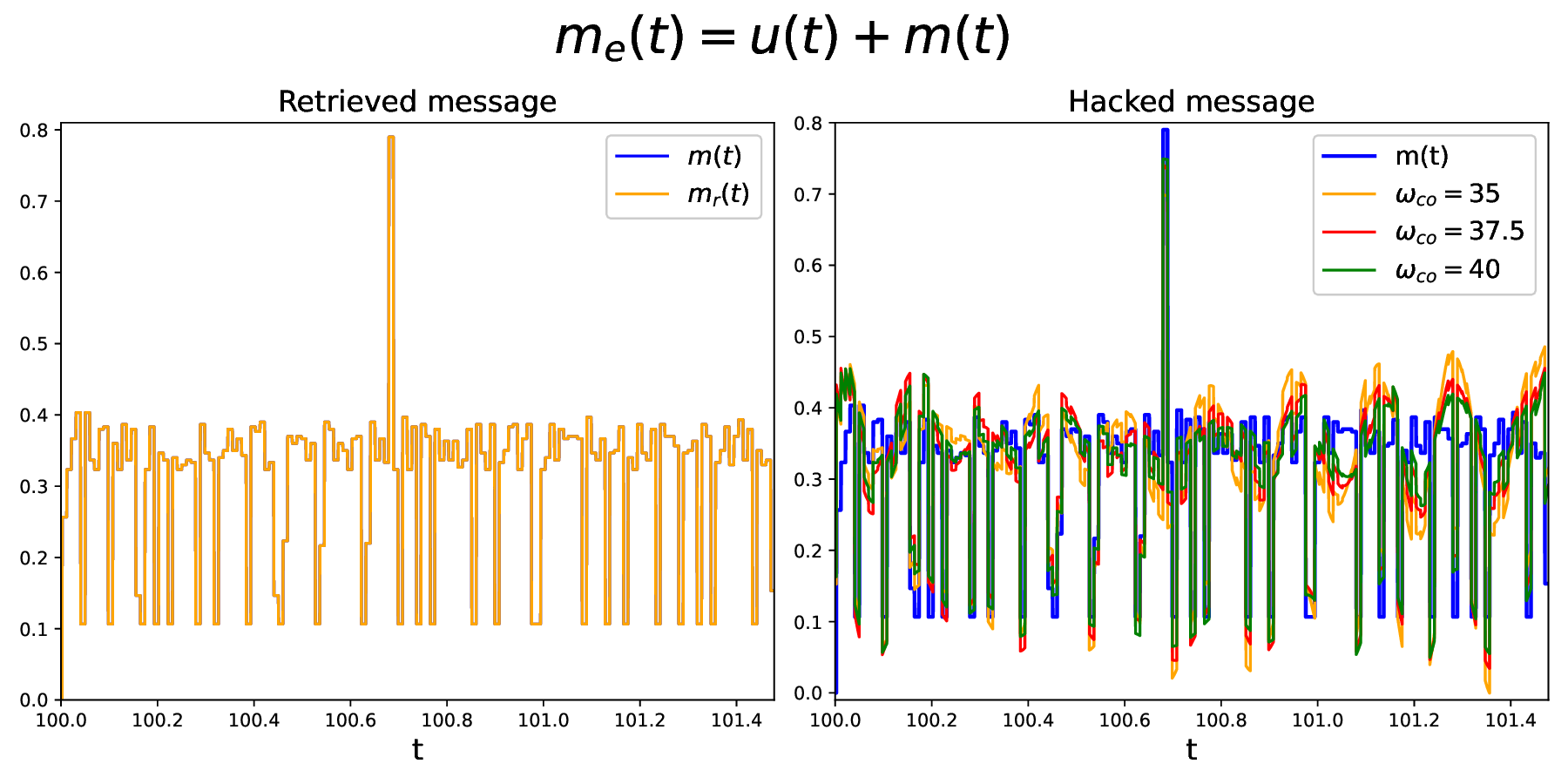}
    \includegraphics[width=8cm]{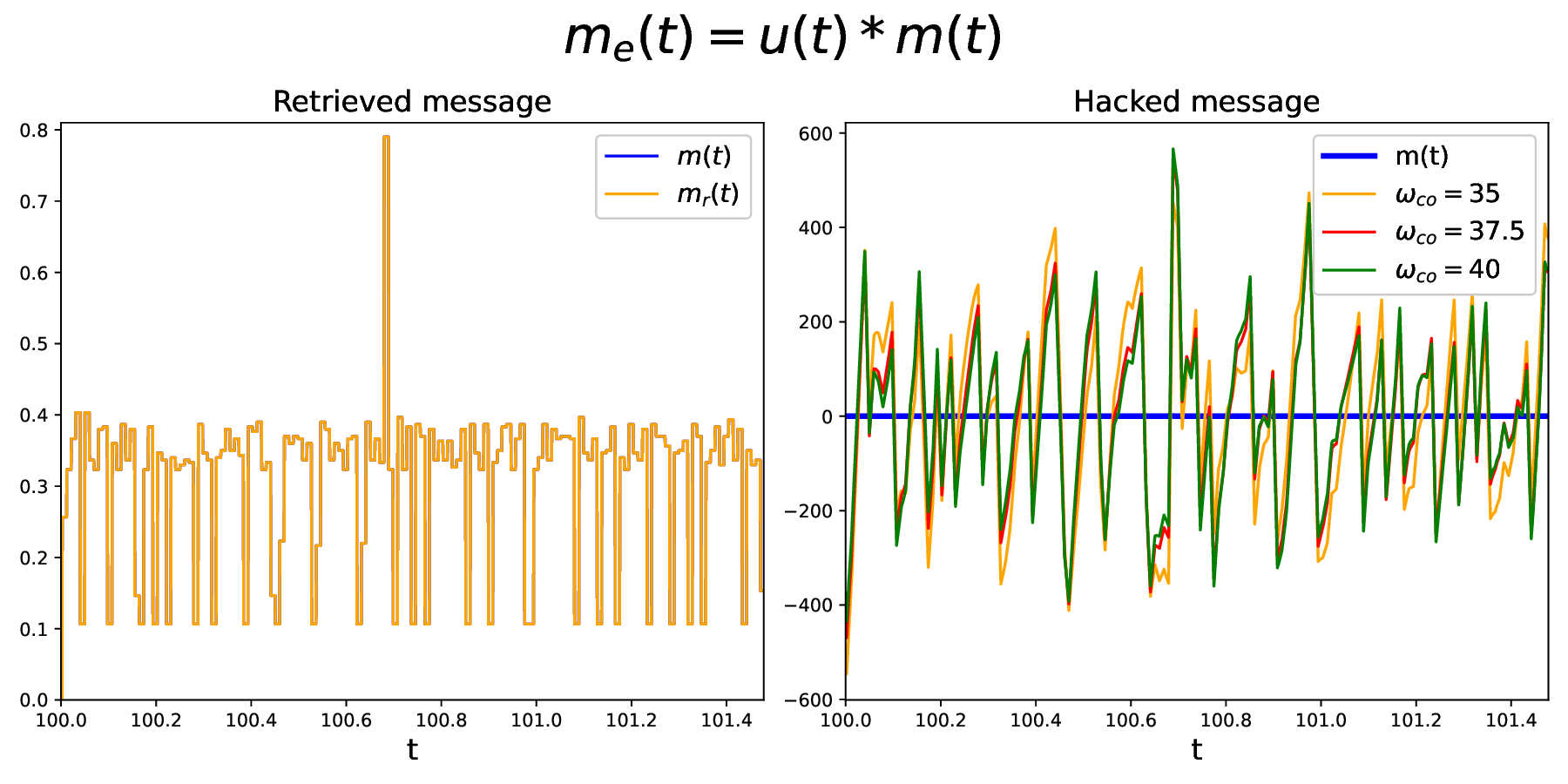}
    \caption{Retrieved message $m_r(t)$ and hacked message $m_h(t)$ using various cut-off frequencies using the addition encryption function Eq. \eqref{eq:EF_sum} (top) and convoEq.  \eqref{eq:EF_conv} (bottom). The message sent is the same text used in subsection \ref{subsection:text_sync}  with the lapse between characters $n=100$ and $t_0 = 100$.}
    \label{fig:Text_conv_100}
\end{figure}

The message is retrieved by the receiver with the $\mathbf{key}_c$ and the inverse encryption functions (\ref{eq:EF_inv_sum}) and (\ref{eq:EF_inv_conv}). The retrieved messages $m_r(t)$ are shown at the left of Figure \ref{fig:Text_conv_100} for the two Encryption Functions.  The PCE has 100\% fidelity for the text message, in the sense that there are no errors in any character of the text.

{\it Hacking the message.} The upper-right graph of Figure \ref{fig:Text_conv_100} presents the results of the hacked message $m_h(t)$ obtained using method \eqref{eq:m_hacked} with different values of the cutoff frequency. The results show at least $95$\% error across the tested frequencies, as only a few characters of the original message were correctly recovered.

For the Encryption Function with convolution \eqref{eq:EF_conv} we do not recover any character correctly when attempting to hack the message, even some parts of the hacked message $m_h(t)$ are negative as seen in bottom right graph Figure \ref{fig:Text_conv_100}.  These results correspond to $n=100$, which represents a low frequency message and is therefore more difficult to hack by inspecting the FT. As previously discussed, for $n=1$, the results in terms of quality of $m_r(t)$ and the hackability of $m_h(t)$ are similar. Given the security is poor for $n = 1$, it is not the ideal choice for the value of $n$.

\section{Conclusions}
\label{sec:conclusions}

In this work, message transmission and encryption using chaotic systems were revisited, comparing the performance of synchronization-based methods with PCE. The study focuses on the reliability of message recovery and the susceptibility of these methods to attacks using Fourier analysis.

For synchronization-based encryption using the Lorenz system, a trade-off was observed between message fidelity and vulnerability. Low-frequency monochromatic messages were effectively immerse within the chaotic signal, as reflected in the Fourier spectrum, but exhibited poor recovery fidelity. In contrast, high-frequency messages were more easily recovered, but introduced identifiable artifacts in the frequency domain, making them susceptible to hacking. When transmitting text messages, synchronization techniques displayed significant degradation in recovery fidelity, particularly as the time step between characters increases, leading to incomplete and errors in the message retrieval.

The Fourier-based hacking method proved highly effective in identifying and partially reconstructing messages. In cases where synchronization failed to recover the message accurately, hacking techniques often outperformed the synchronized receiver, exposing a fundamental weakness of the synchronization based method.

To address these vulnerabilities, Plain Convolution Encryption was introduced using both addition and convolution as encryption functions. The addition-based approach improved message recovery fidelity but remained susceptible to hacking, especially for high-frequency signals. In contrast, the convolution-based method substantially enhanced both message fidelity and resilience to hacking attempts. 
While monochromatic messages encrypted with convolution are successfully recovered, the hacking process cannot recover the original amplitude.For text messages, the convolution-based method achieved 100\% recovery fidelity and proved resistant to hacking using the  Fourier analysis, even in low-frequency transmissions.

In conclusion, the results emphasize the limitations of synchronization-based encryption for secure communication and highlight the advantages of PCE, particularly when employing convolution-based encryption. This method provides a considerable improvement in both message fidelity and security. Future advancements in encryption strategies will be required to address the vulnerabilities exposed by spectral analysis techniques and further enhance the robustness of chaos-based encryption systems.

\appendix 
\section{Converting a List of Characters to a Vector of Integers} \label{appendix:convertir_unicode}

Suppose we have a string of characters that we want to encrypt. Each character can be identified with an integer according to the Unicode (Decimal) format, as shown in Table \ref{Tab: LettersToAscii} for an example alphabet. We can convert the string into a list of integers, which acts like the discrete message $m_i = m(t_i)$ for some values of $t_i$. For example, if the string is "Hello World," the discrete message is $m = \{72, 101, 108, 108, 111, 32, 87, 111, 114, 108, 100\}$.

\begin{table} 
\centering
\begin{tabular}{|c|c|c|c|}
\hline
\textbf{Capital letter} & \textbf{Unicode} & \textbf{Lowercase letter} & \textbf{Unicode} \\
\hline
A & 65 & a & 97 \\
B & 66 & b & 98 \\
C & 67 & c & 99 \\
D & 68 & d & 100 \\
E & 69 & e & 101 \\
F & 70 & f & 102 \\
G & 71 & g & 103 \\
H & 72 & h & 104 \\
I & 73 & i & 105 \\
J & 74 & j & 106 \\
K & 75 & k & 107 \\
L & 76 & l & 108 \\
M & 77 & m & 109 \\
N & 78 & n & 110 \\
O & 79 & o & 111 \\
P & 80 & p & 112 \\
Q & 81 & q & 113 \\
R & 82 & r & 114 \\
S & 83 & s & 115 \\
T & 84 & t & 116 \\
U & 85 & u & 117 \\
V & 86 & v & 118 \\
W & 87 & w & 119 \\
X & 88 & x & 120 \\
Y & 89 & y & 121 \\
Z & 90 & z & 122 \\
\hline
\textvisiblespace & 32 & & \\
\hline
\end{tabular}
\caption{Letters of the alphabet and space with their Decimal-Unicode values. The full catalogue including numbers and other characters can be found at \cite{unicode2024}. }
\label{Tab: LettersToAscii}
\end{table}


\section*{Acknowledgments}
This research is supported by grants CIC-UMSNH-4.9 and Laboratorio Nacional de C\'omputo de Alto Desempe\~no Grant No. 1-2024.


\bibliography{bibliography}

\end{document}